\documentclass[pre,aps,amssymb,amsmath,superscriptaddress,showpacs,
floatfix,nofootinbib,a4paper]{revtex4}

\usepackage{graphicx}
\newcommand{\CM}{{\mathbb C}}

\newcommand{\TM}{{\mathbb T}}
\newcommand{\ZM}{{\mathbb Z}}

\newcommand{\bro}{{\beta _0}}
\newcommand{\ide}{{\bf  \hat I}}
\newcommand{\soppp}{{\bf  \hat S}}
\newcommand{\fib}{{\bf \hat A}}

\newcommand{\fibv}{{\bf  \hat V}}
\newcommand{\fibg}{{\bf \hat G}}

\newcommand{\hres}{{\bf  \hat H}^{\rm res}}

\newcommand{\idb}{{\bf {\hat I}}}
\newcommand{\ssb}{{\bf{\hat S}}}
\newcommand{\hress}{\hat {H}^{\rm res}}

\newcommand{\vt}{{(\vartheta)}}
\newcommand{\vtk}{{(\vartheta,k)}}
\newcommand{\fft}{{\bf  \hat F}}

\newcommand{\aap}{{m_p}}

\begin{document}
\draft
\title{Spinor Dynamics of Quantum Accelerator Modes near Higher Order Resonances.}

\author{Laura Rebuzzini}
\email{laura.rebuzzini@uninsubria.it}
\affiliation{Center for Nonlinear and Complex Systems and
Dipartimento di Fisica e Matematica, Universit\`a dell'Insubria, 
Via Valleggio 11, 22100 Como, Italy.}
\affiliation{Istituto Nazionale di Fisica Nucleare, Sezione di Pavia, 
Via Ugo Bassi 6, 27100 Pavia, Italy.}
\author{Italo Guarneri}
\affiliation{Center for Nonlinear and Complex Systems and
Dipartimento di Fisica e Matematica, Universit\`a
dell'Insubria, Via Valleggio 11, 22100 Como, Italy.}
\affiliation{Istituto Nazionale di Fisica Nucleare, 
Sezione di Pavia, Via Ugo Bassi 6, 
27100 Pavia, Italy.}
\affiliation{CNISM, 
Unit\`a di Como, Via Valleggio 11, 22100 Como, Italy.}
\author{Roberto Artuso}
\affiliation{Center for Nonlinear and Complex Systems and
Dipartimento di Fisica e Matematica, Universit\`a
dell'Insubria, Via Valleggio 11, 22100 Como, Italy.}
\affiliation{CNISM, 
Unit\`a di Como, Via Valleggio 11, 22100 Como, Italy.}
\affiliation{Istituto Nazionale di Fisica Nucleare, Sezione di Milano,
Via Celoria 16, 20133 Milano, Italy.}

\begin{abstract}
Quantum Accelerator Modes were discovered in experiments with Kicked  Cold Atoms in the presence of gravity. They were shown to be  tightly related to resonances of the Quantum Kicked Rotor. In this paper a spinor formalism is developed for the analysis of Modes associated with resonances of 
 arbitrary order 
$q\geq 1$. Decoupling of spin variables from orbital ones is achieved by means of an ansatz of the Born-Oppenheimer type, that generates  $q$ independent  band dynamics. Each of these is described, in classical terms, by a map, and the stable periodic orbits of this map give rise to quantum accelerator modes, which are potentially observable in experiments. The arithmetic organization of such periodic orbits is briefly discussed.

\end{abstract}

\date{\today}


\maketitle

\section{Introduction.}

A kicked system is a Hamiltonian system that 
is periodically driven by pulses of infinitesimal duration.    
More than thirty years after the invention  of the paradigmatic model of such systems, 
namely the Kicked Rotor (KR) \cite{kr},  
Kicked Quantum Dynamics  is still the focus of active research for a twofold reason. 
On the one hand, it has given birth to an ever increasing list of variants of the basic original prototypes, which have provided formally simple models for the investigation of   quantum-classical correspondence and of  some general properties of quantum transport. These include dynamical localization \cite{f3}, anomalous diffusion \cite{khar2,khar1,khar3,gubo},  decay from stable phase-space islands \cite{bkm,bkls,SFGR05}, 
electronic conduction in mesoscopic devices \cite{fbkkrhb,okg,bgr}, 
nondispersive wave packet dynamics \cite{dlryb}, effects of dissipation on quantum dynamics 
\cite{gabriel}, and lately directed transport \cite{TM1,mont,dasum}.   
On the other hand, renewed interest on the physical side has been stimulated by experimental 
realizations \cite{raizen1b,c,KRexp3,phill}, which are now possible,  under  
excellent control conditions,  thanks to the science and technology of cold and ultra-cold atoms. Unexpected  advances of the theory have been  prompted  
by such experiments. For instance, the so-called Quantum Accelerator Modes (QAM) were  discovered in experiments with cold atoms in periodically pulsed optical lattices 
\cite{Ox991,Ox992,Ox995,Ox994}. Their underlying theoretical model is
 a variant of the Kicked Rotor model, the difference being that, in between kicks, 
 atoms are subject to gravity. 
When the kicking period is close to a half-integer multiple of the Talbot time \cite{BB}, which is a natural time scale for the system, a fraction of atoms steadily accelerates away from the 
bulk of the atomic cloud, at a rate and in a direction which depend on various parameter values.
Though QAMs are a somewhat particular phenomenon, their theory 
\cite{FGR022,FGR021,GRF05,Rebk} is a vast repertory of classic items of classical and quantum  mechanics. QAMs are rooted in subtle aspects of the Bloch theory, and have a relation to the Wannier-Stark resonances of Solid-State physics \cite{SFGR05}. They are a purely quantal effect, and yet  they are explained in terms of trajectories  of certain classical  dynamical systems, by means of a ``pseudo-quasi classical"  approximation,  where the role of the Planck constant is played by a parameter $\epsilon$, which measures the detuning of the kicking period from a half-integer multiple of the Talbot time.   This theory hinges on existence of a ``pseudoclassical limit" for $\epsilon\to 0$. That means, for kicking periods close to half-integer multiples of the Talbot time, the quantum dynamics may formally be obtained from quantization of a classical dynamical system, using $\epsilon$ as the Planck constant. This system is totally unrelated from the classical system that is obtained in the proper classical limit
$\hbar\to 0$.\\
 
Experimental and theoretical investigation on QAMs are currently focused on novel 
 research lines: the observation of QAMs in a Bose-Einstein 
condensate \cite{S,rafg,hsgc08}, which allows a precise control on the initial momentum 
distribution, the analysis of QAMs for special values of the 
physical parameters \cite{lemarie,shgc08} and, in particular, 
when the  kicking period is close to a rational multiple 
of Talbot time \cite{GS,GR08}.  
Theoretical aspects concerning the latter problem are considered in the 
present paper.
 \\

 %%
%\begin{figure}
\begin{figure*}
  \includegraphics[width=16cm,angle=0]{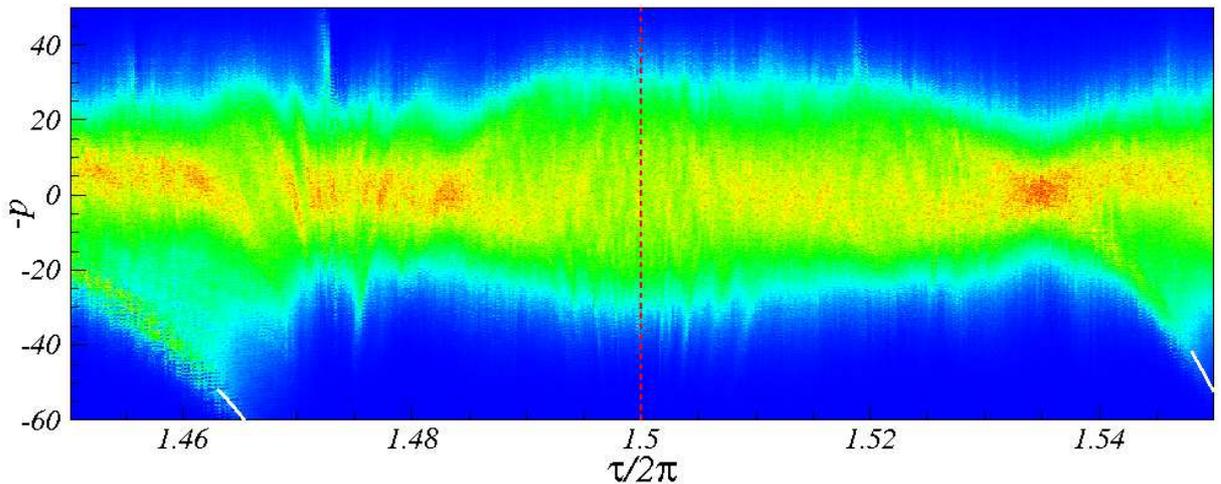}
  \caption{ Momentum distributions, in the time dependent gauge, after
$t=100$ kicks, for different values of the kicking period near the resonance
$\tau_{res} =3\pi$. Red color corresponds to highest probability.
The vertical dashed line corresponds to the resonant value.
 White full lines show the theoretical curves (\ref{acc}), with:
(left) $q=2, (r,s)=(1,1)$ and (right) $q=7, (r,s)=(4,1)$ close to the
 higher order resonance $\tau ^{\rm res}/2\pi =p/q=11/7$. The initial quantum distribution is a gaussian 
wave packet, reproducing the experimental conditions.
The other parameters are: $k=1$ and $g=0.0386$. All our numerical simulation refer to the 
choice $V(\theta)=k \cos \theta$.}
\label{pac-tot}
%\end{figure}
\end{figure*}

QAMS are  connected with  an important feature of the KR model, namely, the KR resonances 
\cite{IzShep1}, which occur whenever the kicking period is rationally related to the internal frequencies of the free rotor.  The dynamics of the rotor at a quantum resonance is invariant under momentum 
translations by multiples of an integer number. The least positive integer $q$, such 
that translation invariance in momentum space 
holds, is the ``order" of the resonance  (sect. \ref{back}). 
The half-Talbot time in atom-optics experiments is 
the period of the KR resonances of order $q=1$ (i.e. ``principal" resonances), 
so the originally observed QAMs are related to KR 
resonances of order $1$.

In this paper we consider quantum motion in the vicinity of a higher order KR resonance 
($q >1$) in the presence of gravity. 
Numerical 
(see fig.\ref{pac-tot}) and heuristic indications \cite{GS} suggest that higher-order KR resonances, 
too,  may give rise to QAMs. This 
has been  substantiated by a theory  \cite{GR08} based on a nontrivial reformulation  of the original 
pseudo-classical approximation. It has been remarked that, in the case of higher-order resonances,  
no pseudoclassical limit exists and similarity to the case of quasi-classical analysis 
for particles with spin was noted, but not explored. 
About the latter general problem \cite{LF91,LF92}, it is known that, although 
no single well-defined classical limit exists, and so no global quasi-classical phase-space approximation in terms of a unique classical Hamiltonian flow is possible,
{\it local} quasi-classical approximations are nevertheless still possible, as
 provided by  bundles of trajectories which 
belong  to a number of different Hamiltonian systems.

In this paper we develop a formulation of the 
  problem of QAMs near higher-order resonances in spinor terms. 
  The quantum evolution at exact  resonance is described by a multi-component wave function, that  is 
   by a spinor of rank $q$, given by the order of the resonance \cite{IzShep1,CaGua}, and is 
  generated by a time-independent  spinor Hamiltonian 
\cite{sokzhicas,SZAC1}.   We show that 
 the small-$\epsilon$ analysis of quantum dynamics is formally 
 equivalent to semiclassical approximation for a particle 
  with spin-orbit coupling.  
 Thus QAMs near higher order resonances constitute a particular, though 
  experimentally relevant, model system, in which this crucial theoretical issue can be explored. 
   The semiclassical theory in
\cite{LF92} is not directly applicable here, because the dynamics is not specified 
by a self adjoint spinor Hamiltonian, but by a spinor unitary propagator instead.   
 We therefore resort to an ``adiabatic" ansatz, which allows  decoupling spin dynamics  from orbital motion. In this way we obtain $q$ distinct and independent orbital one-period propagators.
Each of them may be viewed as the quantization of a formally classical dynamical system, given by a map; however, the ``pseudo-Planck constant" $\epsilon$ explicitly appears in such maps, in a form that precludes existence of a $\epsilon\to 0$ limit for the maps themselves, except for 
the $q=1$ case, in which the pseudo-classical 
 theory of ref. \cite{FGR021,FGR022} is recovered.\\
QAMs, detected by numerical simulations of the exact quantum dynamics near higher order resonances,  tightly  correspond  to stable periodic orbits of the maps.  The acceleration of the modes 
is expressed in terms of the winding numbers of the corresponding orbits and of the order 
of the resonance. Moreover, we derive some 
theoretical results, which generalize those obtained in  \cite{FGR021,GRF05}  
for the principal resonances: a formula for the special values 
of quasi-momenta, which dominate the mode, and a classification of detectable 
modes by a Farey tree 
contruction \cite{far816, HW79}, as a function 
of the gravity acceleration. 

 The paper is organized as follows. In sect.\ref{back} the Floquet operator, 
describing one-step evolution of a kicked atom in a free-falling frame, 
is recalled and the resonant spinor dynamics in Kicked Particle (KP) model 
is briefly reviewed; in sect.\ref{sodec}, 
the quantum motion in the vicinity of a resonance of arbitrary order is related 
to the problem of a particle with spin-orbit coupling. 
 In sect.\ref{map}, a ``formally" classical description of the orbital dynamics, associated 
 to the QAMs, is achieved.  
Finally, in sect.\ref{exper} connections between  
the theoretical results and possible experimental findings are discussed.

\section{Background.}
\label{back}

\subsection{Floquet operator in the ``temporal gauge".}

In the laboratory frame, the quantum dynamics of the atoms moving under the joint action 
of gravity and of the kicking potential, is ruled by the 
time-dependent Hamiltonian (expressed in dimensionless units):
\begin{equation}
\label{ham-l}
\hat H_L(t)=\frac {\hat p^2}{2}-\frac {\eta}{\tau}\hat x+ k V (\hat x)
\sum_{n=-\infty}^{+\infty} \delta (t-n\tau ),
\end{equation}
where $\hat p$ and $\hat x$ are the momentum and position operators.  
The potential $V(x)$ is a smooth periodic function of spatial period $2\pi$. Denoting $M,T, {\mathrm K}, {\mathrm g}$ and $2\pi/G$  the atomic mass, the temporal period of the kicking, the kicking strength, the gravity acceleration and the spatial period of the kicks, respectively, the momentum, position and mass of the atom in (\ref{ham-l})
are rescaled in units of $\hbar G$, $G^{-1}$ and $M$.
 The three dimensionless parameters $k,\tau$ and
$\eta$ in (\ref{ham-l}), which fully characterize the dynamics, 
are expressed in terms of physical quantities by $k={\mathrm K}/\hbar$,
$\eta =M{\mathrm g}T/(\hbar G)$ and $\tau =\hbar TG^2 /M=4\pi T/T_B$. 
$T_B=4 \pi M/(\hbar G^2)$ is the Talbot time \cite{BB} and $g=\eta /\tau$ is the rescaled 
gravity acceleration.  
 Throughout the following $\hbar=1$ is understood.

For $\eta=0$, 
the Hamiltonian (\ref{ham-l}) reduces to that of the Kicked Particle model (KP), 
which is a well-known variant of the Kicked Rotor model (KR),  corresponding 
to the particular choice: $V(x)=k\cos (x)$.  The KP differs from the KR 
because the eigenvalues of particle momentum are continuos while those of the angular momentum 
of the rotor are discrete. Due to Bloch theorem, the invariance of the KP Hamiltonian,  
under space translations by $2\pi$, implies conservation of the quasi-momentum $\beta$, which, 
in the chosen units, is the fractional part of the momentum. The particle 
 momentum is decomposed as $p=N+\beta$ with 
$N\in {\mathbb Z}$ and $0\leq \beta <1$. Conservation of quasi-momentum 
 enables a Bloch-Wannier fibration of the particle dynamics: 
the particle wave function is obtained by a superposition of Bloch waves, 
describing the states of  independently evolving kicked rotors with different 
values of the quasi-momentum (called $\beta$-rotors).

A remarkable feature of Hamiltonian (\ref{ham-l}) is that, unless rescaled gravity $g=\eta/\tau$ assumes exceptional commensurate values, the linear potential term breaks invariance under $2 \pi$ space translations. Such an invariance may be recovered by going to a temporal gauge, where momentum is measured {\em w.r.t.} free fall. This transformation gets rid of the linear term and the new Hamiltonian reads  \cite{FGR021}:
\begin{equation}
\label{ham-ff}
\hat H_g (t)=\frac 12 (\hat N +\beta +\frac \eta\tau t)^2+k  V(\hat \theta)
\sum_{n=-\infty}^{+\infty} \delta (t-n\tau ).
\end{equation}
where $\theta =x\; {\rm mod}(2\pi)$, $\hat{N}=-id/d\theta$ with 
periodic boundary conditions.

 The quantum motion of a $\beta$-rotor in the ``temporal gauge" (that is,  ``in the falling frame") is described by the following  Floquet operator  
on $L^2 ({\mathbb T})$ (  ${\mathbb T}$ denotes the 1-torus, parametrized by $\theta\in [-\pi,\pi [$): 
\begin{equation}
\label{fo}
\hat U_\beta (n) =e^{-ikV(\hat \theta )}e^{-i\frac {\tau}{2}(\hat N +\beta +\eta n+\frac {\eta}{2})^2}.
\end{equation}
where $n\in\ZM$ denotes the number of kicks.  
The operator (\ref{fo}) describes evolution from time $t=n\tau$ to time
$t=(n+1)\tau$. 

\subsection{Quantum Resonances.}
%\label{qrr}

 We consider the problem of Quantum Accelerator Modes in the 
vicinity of a generic resonance of the $\beta$-rotor. 
The concept of quantum resonance (QR) is reviewed in this subsection. 

A QR occurs whenever quantum evolution  commutes with a nontrivial group of momentum translations. A momentum  translation $\hat{N}\to\hat{N}+\ell$ (recall $\hbar=1$) with $\ell\in\ZM$ is described by
the operator $\hat T^{\ell}=e^{i{\ell}\hat\theta}$. In the following we assume $\eta=0$ and then  the  operator (\ref{fo}) is time-independent. It commutes with  $\hat{T}^{\ell}$ if and only if \cite{dd}:
$i)$ $\tau  /2\pi =p/q$ with $p,q$ coprime integers;
$ii)$ ${\ell}=rq$ with $r\in {\mathbb N}$; $iii)$ $\beta= \nu/rp +rq/2$(mod 1), with $\nu\in{\mathbb Z}$.
In this paper we
restrict to ``primary" resonances, i.e. to resonances with $r=1$ and $\ell =q$; in this case,
$q$ defines the order of the resonance. QR of order 1 are called ``principal resonances". 

 A theory for QAMs in the vicinity of 
principal resonances was proposed in \cite{FGR022,FGR021}. In this paper we consider 
 quantum resonances of arbitrary order $q\geq 1$. The resonant values 
 of the kicking period $(i)$, expressed in physical units, coincide with rational multiples of half of the Talbot time.
 
We generically denote $\hat{U}_{\mbox{\tiny res}}$ the operator (\ref{fo}) at resonance, and $\beta_0$ the resonant values of quasi-momentum, given by the above condition $(iii)$. 

\subsection{Bloch theory and spinors.}
%\label{bthspin}

Translation invariance under $\hat{T}^q$ enforces conservation of the Bloch phase $\xi\equiv\theta$ mod $2\pi/q$,
taking values in the Brillouin zone $\mathbb B=[-\pi /q, \pi/q[$. Loosely speaking, this means that
$\theta$ only changes by multiples of $2\pi/q$, so $\xi$ has the meaning of  ``quasi-position".
 As we show below, a Bloch-Wannier fibration of the rotor dynamics holds 
with respect to the quasi-position $\xi$, at a QR.

We use a rescaled  quasi-position $\vartheta\equiv q\xi$, and accordingly resize the Brillouin zone
to $[-\pi,\pi[$. In all representations where quasi-position is diagonal, the state $|\psi\rangle$ of the rotor
is described by a
$q$-spinor ${\phi}$, specified by $q$ complex functions
$\phi_l(\vartheta)=\langle \vartheta,l |\psi\rangle$, $(l=1,\ldots,q) $.
We shall use a  representation where the spinor $\pmb{\phi}(\vartheta)$,
which corresponds to a given rotor wavefunction $\psi(\theta)=\langle\theta|\psi\rangle$,  is defined by:
\begin{equation}
\label{scomp2}
 \phi _l (\vartheta )
\;=\;\frac {1}{\sqrt {2\pi}}\;\sum _{m\in {\mathbb Z}} \hat
\psi (l+mq) e^{im\vartheta}
\quad\quad\quad l=0,...,q-1.
\end{equation}
where $\hat \psi(n)$, $n\in\ZM$  are the Fourier coefficients
of $\psi (\theta)$. 
Equation (\ref{scomp2}) defines a unitary map $\mathfrak a$ of $L^2(\TM)$ onto
$L^2(\TM)\otimes\CM^q$. Under this map, the (angular) momentum operator $\hat N$ is tranformed to:
\begin{equation}
\label{recipe}
\hat N =-i \partial_\theta \;\to \; {\mathfrak a}(\hat N) {\mathfrak a}^{-1}\; =\;
 -iq\partial_\vartheta
\otimes \ide + \hat I\otimes \soppp,
\end{equation}
where $\hat I$ and $\ide$ are  the identity operators in $L^2({\mathbb T})$ and in $\CM^q$ respectively, and $\soppp$ is the spin operator in $\CM^q$:
\begin{equation}
\label{bthspin}
\soppp =\sum_{l=0}^{q-1} l
 | l \rangle \langle l |.\;
\end{equation}
where $|l\rangle$, $l=1,\ldots,q-1$ is the canonical basis in $\CM^q$.
Thus, in spinor representation,  the momentum operator is  the sum of the
orbital operator $-iq\hat \partial _\vartheta \otimes \idb$
and the spin operator $\hat I \otimes \ssb$.
In this picture, the rotor is characterized by ``orbital" observables
($\vartheta$,  $-i\partial _\vartheta$) and by the spin
observable.

Bold symbols denote vectors in $\CM^q$ and matrices.

\subsection{Resonant spin dynamics.}
%\label{resspin}

At resonance, quasi-position is conserved under the discrete-time evolution defined by (\ref{fo}), so, whenever
it has a definite value $\vartheta$, no ``orbital" motion occurs, and spin alone changes  in time. Therefore   the evolution is described by a unitary $q\times q$ matrix $\fib(\vartheta)$ such that, as
$\psi(\theta)$ evolves into $\hat U_{\rm res}\psi(\theta)$,
the corresponding spinor $\pmb{\phi}(\vartheta)$ evolves into
the spinor $\fib(\vartheta)\pmb{\phi}(\vartheta)$.
The explicit form of the spin propagator $\fib(\vartheta)$ is easily computed by using (\ref{fo}) under resonance conditions. With the specific choice $V(\theta)=\cos(\theta)$, one finds
(details can be found in appendix \ref{exres}): 
%\footnote{inserire la forma esplicita di $\exp(ikV)$.}
\begin{eqnarray}
\label{fibfl}
& & \fib(\vartheta)\;=\;e^{-ik\fibv(\vartheta )}e^{-i\fibg},\\
\label{fibfl1}
& & \fibg\;\equiv\;\fibg _{p,q,\bro }\;=\;\pi \frac pq (\soppp +\bro\ide )^2,\\
\label{fibfl2}
& & \fibv (\vartheta )=
\frac 12 \left\{
\sum _{l=0}^{q-2} \left( | l \rangle \langle l+1 | +
| l+1 \rangle \langle l |\right)
+|0\rangle \langle q-1 | e^{i\vartheta } +|q-1 \rangle \langle 0
| e^{-i\vartheta} \right\}.
\end{eqnarray}

\subsection{Bands.}

\begin{figure}
  \includegraphics[width=8cm,angle=0]{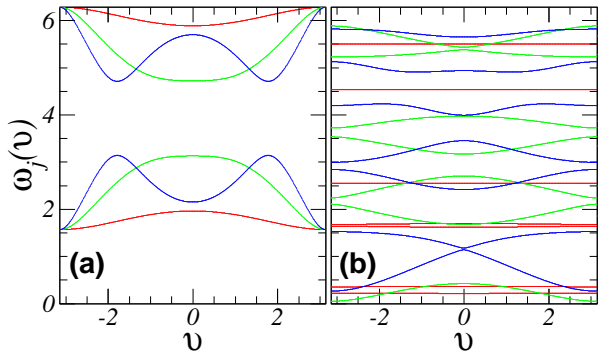}
  \caption{ Eigenvalues of the resonant Hamiltonian $\hres (\vartheta )$ for different values of the kicking constant $k=1$ (red), $3$ (green) and $5$ (blue) and (a) $q=2, p=3$, (b) $q=7, p=11$. }
\label{levels}
\end{figure}

The  ``resonant Hamiltonian"
$\hres (\vartheta)$ is a hermitean matrix  of rank $q$ such that:
\begin{equation}
\label{fibhres}
 \fib(\vartheta )=e^{-i\hres (\vartheta)}.\\
\end{equation}
It is uniquely defined, under the condition that its eigenvalues (\emph{i.e.},  the eigenphases of
$\fib(\vartheta)$) lie in $[0,2\pi[$.
Explicit calculation of eigenvalues and eigenvectors of
$\hat{\bf A}(\vartheta)$, hence of the resonant Hamiltonian,  is trivial for $q=1$, and
is easily performed for $q=2$ in terms of Pauli matrices  \cite{SZAC1}  (such a
 case is reviewed in appendix \ref{appq2}). However, for $q>2$ analytical calculation is prohibitive.\\
Eigenphases of $\hat{\bf A}(\vartheta)$ are
 smooth periodic function of quasi-position $\vartheta$.
 As $\vartheta$ varies in $[-\pi,\pi[$, they sweep bands in the quasi-energy spectrum of the resonant evolution described by $\hat U_{\mbox{\tiny res}}$ \cite{cs}.
 They also depend  on the kicking strength $k$ and will be denoted by $\omega_l=\omega _l(\vartheta, k)$ in the following ($l=0,...,q-1$).
 In the case $q=1$,  $\omega _{0}(\vartheta, k)=k\cos (\vartheta)$.
 For $q>1$ the eigenvalues  are
 nontrivial functions of the kick strength $k$.  For fixed $q >2$ bandwidths tend to increase with $k$, eventually giving rise to complex patterns of avoided crossings.
 Examples of $\vartheta$- and $k$-dependence of eigenphases
 are shown in fig.\ref{levels} for $q=2$ (a) and $q=7$ (b). For $q>2$ the bandwidths depend also 
 on $l$. 
 
 In the resonant representation (i.e. in the representation in which the resonant  
 propagators (\ref{fibhres}) are diagonal), the spinor components (\ref{scomp2}) evolve 
 independently.

%%%%%
\section{Near-resonant dynamics  and spin-orbital decoupling.}
\label{sodec}

We are interested in quantum motion, described by (\ref{fo}), in the vicinity of a
QR, namely when the kicking period is $\tau=2\pi p/q +\epsilon$, where the detuning $\epsilon$ of the period from the resonant period  $\tau _{\rm res}=2\pi p/q$ is assumed to be small.
  The one-step evolution operator (\ref{fo}) may be factorized as:
\begin{eqnarray}
\label{fodec0}
& & \hat U_\beta (n) =\hat U_{\rm res}\cdot
\hat U_{\rm nr} (n)\\
\label{fodec1}
& &  \hat U_{\rm nr}(n)=
e^{-i\left[ \frac {1}{2}\epsilon\hat N^2+ D_n  \hat N\right]},
\end{eqnarray}
where $D_n =\tau (\beta + \eta n +{\eta}/{2})-2\pi p \bro/q.$
%\begin{equation}
%\label{ddt}
%D_n =\tau (\beta + \eta n +{\eta}/{2})-2\pi p \bro/q.
%%=\tau(\eta n+\eta/2)+2\pi p(\beta-\beta_0)/q+\epsilon\beta
%\end{equation}

\subsection{``Adiabatic" decoupling of spinor and orbital motions.}  

Translation invariance (in momentum) is now broken  by  $\hat U_{\rm nr}$ , 
so quasi-position is not conserved any more.
The evolution of a spinor ${\phi}\in L^2(\TM)\otimes\CM^q$ is ruled by the
time-dependent Schr\"odinger equation:
\begin{eqnarray}
&& i\epsilon \frac {\partial}{\partial t}{\phi }\; =\;
\hat {H} (t ){\phi} \;,
\qquad\qquad\qquad
\hat {H} (t ) =
\epsilon
\hress \cdot \sum _{n=-\infty}^{+\infty}\delta (t-n)+ \hat {H}_0 (t),
\label{ham-sp}
\\
&& \hat {H}_0 (t)=
\frac 12 \epsilon ^2
(-iq\partial_\vartheta
\otimes \ide + \hat I\otimes \soppp)^2 +\epsilon D_{[t]} (-iq\partial_\vartheta
\otimes \ide + \hat I\otimes \soppp)
\label{ham-sp1}.
\end{eqnarray}
where $[t]$ denotes the integer part of $t$ and $(\hress \phi) (\vartheta)=
\hres (\vartheta) \pmb{\phi}
(\vartheta)$. 
Note that $\hat {H}_0 (t)$ is constant in between kicks. 
  
Both sides of the Schr\"odinger equation have been multiplied by $\epsilon$,
to make it apparent that the detuning $\epsilon$ plays the role of an effective Planck constant
in what concerns the motion between the $\delta$-kicks. \\

The spinor components are mixed by (\ref{ham-sp1}) during the evolution;  
we use an ansatz of  Born-Oppenheimer type in order to decouple orbital (slow) motion  from spin (fast) motion. 

The detuning $\epsilon$ controls as well the separation between the different time scales 
of the system. At exact resonance (i.e. $\epsilon=0$) the decoupling is exact, because motion 
is restricted to the eigenspaces of the resonant propagator (\ref{fibhres}). 
 These subspaces are defined by the spectral decomposition of the resonant Hamiltonian, which we
 write in the form:
\begin{equation}
\label{spr}
\hres\vt\;=\;\sum\limits_{j=0}^{q-1}\;\omega_j(\vartheta,k)\;\hat{\bf P}_j(\vartheta)\;\;\;,\;\;\;
\hat{\bf P}_j(\vartheta)\;=\;|\pmb{\varphi}_j(\vartheta )\rangle\langle\pmb{\varphi}_j(\vartheta)|\;,
\end{equation}
where $\pmb{\varphi}_j(\vartheta )$ is the normalized eigenvector of $\hres\vt$ which corresponds
to the eigenvalue $\omega _j \vtk$. For each value of $\vartheta$, the
operators $\hat{\bf P}_j(\vartheta)$ in (\ref{spr}) are projectors in $\CM^q$. We denote 
by $\hat P_j$ the projectors
in the full Hilbert space 
$L^2(\TM)\otimes\CM^q$, which act on spinors according to $(\hat{P}_j{\phi})(\vartheta)=\hat{\bf P}_j(\vartheta)\pmb{\phi}(\vartheta)$. The subspaces ${\cal H}_j$ whereupon the ${\hat P}_j$ project are the ``band subspaces" and are not invariant for the full Hamiltonian (\ref{ham-sp}).  
By using the ansatz that ``band subspaces" be almost invariant for small $\epsilon$, 
we next decouple the (assumedly ``fast") spin variables from the orbital (``slow") ones. 
We assume that the decoupled evolution inside the band subspaces provides a good 
description of the exact evolution when $\epsilon$ is small, because 
the leading error terms are linear in $\epsilon$.

Our  approximation
 consists in replacing the exact dynamics, ruled by the Hamiltonian in (\ref{ham-sp}), 
 (\ref{ham-sp1}) by an ``adiabatic" evolution, generated by the 
Hamiltonian:
\begin{gather}
\label{bo}
\hat{H}^{\mbox{\tiny diag}}(t)\;=\;\sum\limits_{j=0}^{q-1}\;\hat{P}_j\;\hat{H}(t)\;\hat{P}_j\;=\;
\epsilon \hress \cdot \sum _{n=-\infty}^{+\infty}\delta (t-n)\;
+\;\sum\limits_{j=0}^{q-1}\;\hat{P}_j\;\hat{H}_0(t)\;\hat{P}_j\;.
\end{gather}

In the case of time-independent Hamiltonians, such projection on ``band subspaces", aimed at separating fast and slow time scales, is basically a Born-Oppenheimer approximation \cite{adth}. 
In the case of kicked dynamics this projection  should be performed  on the ``effective", time-independent  Hamiltonian ${\hat H}_{\mbox{\tiny eff}}$, which
generates over a unit time the same evolution as does the kicked Hamiltonian. The
 effective Hamiltonian is not known in closed form, although it can be expressed by a sum 
of infinite terms, ordered in powers of $\epsilon$ \cite{SZAC1,danaar}. 
Our ansatz is somehow related to a rough approximation ${\hat H}_{\mbox{\tiny eff}}\simeq
\epsilon \hress +{\hat H}_0$. 
 We assume this is valid in some restricted parameter regimes (see further comments in Sect. \ref{map}).
\\

A spinor in ${\cal H}_j$ has the form $\psi(\vartheta)\pmb{\varphi}_j(\vartheta)$ with $\psi\in L^2(\TM)$ and may thus be described by a scalar wavefunction  $\psi(\vartheta)$ (the amplitude of the spinor on the $j$-th resonant eigenstate).  Evolution inside the band subspace ${\cal H}_j$ is ruled by the ``band Hamiltonian" $\hat{H}_j(t)=\hat{P}_j\;\hat{H}(t)\;\hat{P}_j$ and
 direct calculation by using (\ref{ham-sp}) shows that band Hamiltonians have the following form:
\begin{gather}
\label{bo1}
{\hat H}_j(t)\;=\;\;\epsilon\omega_j(\vartheta,k)\sum\limits_{t'=-\infty}^{+\infty}
\delta(t-t')\;+\;\hat{H}_0^{(j)}(t)\;,
\end{gather}
where:
\begin{eqnarray}
\label{matrel1}
&& \hat{H}^{(j)}_0(t)\; = \;-\frac 12 \epsilon ^2 q^2 \partial _\vartheta ^2 -
\left (\epsilon ^2 q^2 \langle \pmb{\varphi}_j | { \pmb{\dot  \varphi}_j}  \rangle +i\epsilon ^2 q S_j +i \epsilon
q D _{[t]}\right)\partial _\vartheta +\frac 12 \epsilon ^2 \left( S^{''}_j
-q^2  \langle \pmb{\varphi}_j | {\pmb{\ddot \varphi}_j}  \rangle
-i 2q S^{'}_{l}\right) +\nonumber \\
 &&  \qquad\qquad \qquad\qquad  +\epsilon D _{[t]} \left( S_j-iq  \langle
 \pmb{\varphi}_j |  {\pmb{\dot \varphi}_j}  \rangle \right),
\end{eqnarray}
where dots denote derivatives with respect to $\vartheta$, and
\begin{equation}
\label{ssss}
S_j\vt =\langle \pmb{\varphi}_j \vt | {\hat {\bf S}}|  \pmb{\varphi}_j \vt  \rangle\;, \quad
S^{'}_j\vt =\langle \pmb{\varphi}_j \vt | {\hat {\bf S}}| \pmb {\dot {\varphi}} _j \vt  \rangle, \quad
\quad
S^{''}_j\vt =\langle \pmb{\varphi}_j \vt | {\hat {\bf S}^2}|  \pmb{\varphi}_j \vt  \rangle\;.
\end{equation}

\subsection{Band Hamiltonians.}

We now note that the problem can be formulated as the evolution of a particle in a fictitious magnetic 
field, which takes into account the average effects of spin degree of freedom on the orbital motion.

We derive a simpler form for the band Hamiltonians (eqs.(\ref{eqsc-eff})). 
 By the introduction of magnetic vector  
and scalar potentials, the operator (\ref{matrel1})
may be written  in  the form:
\begin{equation}
\label{h1-pot}
\hat {H}_j(t)\;=\;\frac 12 \epsilon ^2 q^2
\left(-i\partial _\vartheta -{\mathcal A}_j\vt\right)^2 +\epsilon
q D _{[t]}
\left(-i\partial _\vartheta -{\mathcal A}_j\vt\right) +\frac 12 \epsilon ^2{\mathcal B}_j\vt\;.
\end{equation}
The ``geometric" vector potential ${\mathcal A}_j \vt$ and the scalar potential ${\mathcal B}_j\vt$  are determined
by the structure of the resonant eigenvectors $\pmb{\varphi}_j(\vartheta)$, via the following  relations:
\begin{eqnarray}
\label{ab1}
&& {\mathcal A}_j \vt =i \langle \pmb{\varphi}_j \vt |{\pmb{\dot\varphi}}_j\vt \rangle -\frac 1q S_j \vt\\
\label{ab2}
&& {\mathcal B}_j\vt = S^{''}_j \vt +2q\; \Im S^{'}_j \vt -q^2  {\mathcal A}_j^2 \vt
+ q^2 \langle {\pmb{\dot\varphi}_j} \vt |{\pmb{ \dot\varphi}}_j \vt \rangle.
\end{eqnarray}
Reality of such potentials
follows from (\ref{ssss}) and from the fact that $\langle \pmb{\varphi}_j \vt | {\pmb{ \dot\varphi }}_j \vt \rangle $ is purely imaginary thanks to normalization.  The vector potential is 
gauge-dependent; 
eigenvectors $\pmb{\varphi}_j \vt$ are determined up to   arbitrary  $\vartheta$-dependent
phase factors and so operator (\ref{h1-pot}) may be further simplified by a gauge transformation,
$
 \pmb{\varphi}_j \vt \to \pmb{\varphi}_j \vt e^{i\lambda _j \vt}.
$
 Under such a transformation,  ${\mathcal A}_j \vt$ changes to
 $\tilde {\mathcal A}_j \vt ={\mathcal A}_j \vt  -\dot {\lambda} _j \vt$, and ${\cal B}_j\vt$ does not change.  The transformation may be chosen so that
 \begin{equation}
 \label{potco}
\tilde{\cal A}_j\vt\;=\;\mbox{\rm const.}\;=\;-\gamma_{j,q}\;-\varsigma\;\equiv \alpha _j\;, 
\end{equation}
where:
$$
\gamma_{j,q}=\frac1{2\pi i}\int_{-\pi}^{\pi}d\vartheta\;\langle\pmb{\varphi}_j\vt|{\pmb{\dot\varphi}}_j\vt\rangle\;\;\;,\;\;
\varsigma=\frac1{2\pi q}\int_{-\pi}^{\pi}d\vartheta\;S_j\vt\;.
$$
This immediately follows from (\ref{ab1}) and from the requirement, that  eigenvectors
 be single-valued.  Note that
$2\pi\gamma_{j,q}$ is the geometric (Berry's) phase \cite{berph,bsimon}.
We thus assume  ${\tilde {\mathcal A}}_j=\alpha _j$:  
this choice corresponds to the Coulomb gauge.\\
In conclusion, in the $j$-th band subspace, the 
%effective 
 band  dynamics is described by the following
Schr\"odinger equation:
\begin{eqnarray}
\label{eqsc-eff}
&& i\epsilon \frac {\partial}{\partial t}\psi(\vartheta ,t)\; =\;
%\hat {H}_j^{\mbox{\tiny eff}}(t)\;\psi(\vartheta ,t)
\hat {H}_j (t)\;\psi(\vartheta ,t),
\nonumber  \\
%&& \hat {H}_j^{\mbox{\tiny eff}}(t)\;=\;
&& \hat {H}_j  (t)\;=\;
\epsilon\omega_j(\vartheta,k)\;\sum _{n=-\infty}^{\infty}\delta (t-n)+
\frac 12 \epsilon ^2 q^2\left(-i\partial_\vartheta -\alpha _j\right)^2
+\frac12\epsilon^2{\mathcal B}_j\vt + \epsilon
q D _{[t]}
\left(-i\partial_\vartheta -\alpha _j \right)
\end{eqnarray}

The multicomponent Schr\"odinger equation  (\ref{ham-sp}), for the $q$-spinor wave function  
$\phi (\vartheta,t)$, is then 
reduced to $q$ scalar Schr\"odinger equations  (\ref{eqsc-eff}), each of which 
determines the independent evolution of a rotor wave function $\psi (\vartheta ,t)$.

%%%%%%%%%

\section{Pseudo-classical description of orbital motion.}
\label{map}

 We now derive a 
 description of the  dynamics of the orbital observables
($\vartheta$,  $-i\partial _\vartheta$), restricted inside each of the band subspaces 
 ${\cal H}_j$,  by ``formally" classical equations of motion.

We introduce a ``pseudo-classical'' momentum operator $\hat I$, defined as follows:
\begin{equation}
\label{clm}
\hat I=-i\epsilon \partial_\vartheta\;,
\end{equation}
which differs from the orbital momentum because of the replacement of 
 the Planck constant ($=1$) by $\epsilon$.
If the same role is granted to $\epsilon$ in eqs.(\ref{eqsc-eff}),  then, in classical terms,
 the effective
band dynamics in the $j$-th band subspace looks like a rotor dynamics,
with angle coordinate $\vartheta$ and conjugate momentum $I$, ruled  by the
kicked Hamiltonian:
\begin{eqnarray}
\label{hamj}
& & H_j (\vartheta , I , t) = \epsilon\omega_j(\vartheta,k)\sum _{t=-\infty}^{+\infty}
\delta (t-t')+F_j (\vartheta, I,t), \nonumber \\
&&
F_j (\vartheta, I,t)=
 \frac 12 q^2  I^2 +D_{[t]} q I -\epsilon q^2 \alpha _j I +
 \frac 12 \epsilon^2 {\mathcal B}_j\vt.
\end{eqnarray}
Terms, independent of $I$ and $\vartheta$, have been neglected. This Hamiltonian describes a classical kicked dynamics. By dropping terms beyond first order
 in $\epsilon$, the map from immediately after the $n$-th kick to immediately after the $(n+1)$-th kick is:
\begin{eqnarray}
\label{map0}
& \vartheta _{n+1} =\vartheta _n + q^2I_n +2\pi\Omega q n +\varrho\;, & \qquad\qquad
{\rm mod} \;  (2\pi), \nonumber\\
%& I_{n+1} =I_n + \epsilon f_j(\vartheta_{n+1},k)\;, &  \qquad \qquad
& I_{n+1} =I_n - \epsilon \dot \omega_j(\vartheta_{n+1},k)\;, &  \qquad \qquad 
\end{eqnarray}
where $\Omega=\eta\tau/(2\pi)$ and 
$\varrho=q(-\epsilon q \alpha _j+\pi \Omega +\tau\beta -2\pi p \beta_0 /q)$.
%$\Omega=\eta\tau/(2\pi)$ (cp. eqn.(\ref{ddt})), and
%\begin{gather}
%\varrho=q(-\epsilon q \alpha _j+\pi \Omega +\tau\beta -2\pi p \beta_0 /q)
%\nonumber\\ f_j(\vartheta,k)=-\partial_{\vartheta}\omega_j(\vartheta,k).
%\end{gather}
%\begin{gather}
%\varrho=q(-\epsilon q \alpha _j+\pi \Omega +\tau\beta -2\pi p \beta_0 /q). 
%\end{gather}

The meaning of  the pseudoclassical map (\ref{map0}) as a description of the nearly resonant quantum dynamics will be discussed in section \ref{meaning}.

\subsection{Pseudo-classical maps and Quantum Accelerator Modes.}
\label{pcmaps}
We now describe how quantum accelerator modes appear in the present framework.

\begin{figure}
  \includegraphics[width=8cm,angle=0]{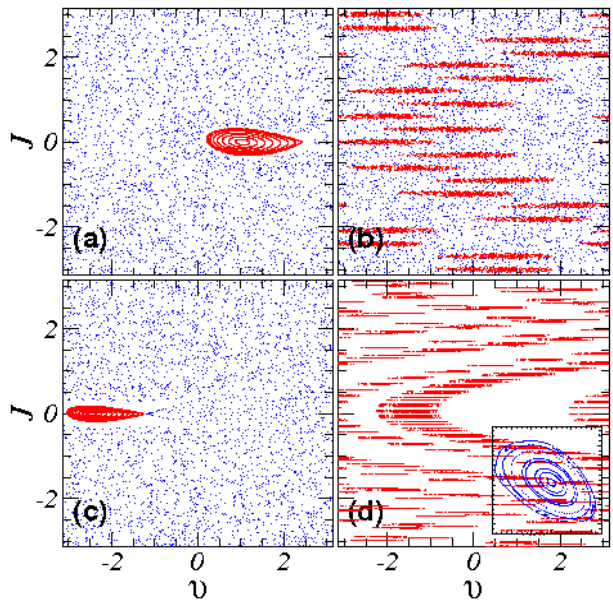}
  \caption{ Phase portraits of maps (\ref{mapq2}) with $k=1$ and $g=0.0386$ and 
  (a) $\tau /2\pi=1.445\; (\epsilon =\tilde k =-0.2827),\; \tau\eta=2\pi\Omega =3.2261, (r,s)=(1,1)$, 
  of map (\ref{mapq2}) with $j=1$; 
  (b) $\tau /2\pi=1.5025\; (\epsilon =\tilde k=0.0157),\; \tau\eta =3.4401,(r,s)=(23,21)$, 
  of map (\ref{mapq2}) with $j=0$; 
   (c) $p/q =11/7, \tau /2\pi=1.5375 \; ( \epsilon =\tilde k =0.2132),\; \tau\eta=3.6023, (r,s)=(4,1)$, of
   map (\ref{mapj}) with $j=3$; 
  (d) $p/q= 22/15, \tau /2\pi=1.485\; (\epsilon =\tilde k = 0.1152),\; \tau\eta =3.3605, r/s=(8,1)$, 
  of map (\ref{mapj}) with $j=15$. 
  The periodic orbit has period $r=132$ and it is associated to 132 stability islands; $r$ and $s$ are not coprime ($r=1056$, $s=132$). In the inset a magnification of one of the small islands of the chain is shown.
 }
\label{phsp-q2p3-q7p11-q15p22}
\end{figure}

The explicit dependence on time of map (\ref{map0}) is removed by changing the momentum variable to
$J_n=q^2 I_n + 2\pi\Omega q n + \varrho$. In the variables $(J,\vartheta)$ the map is $2\pi$-periodic in $J$ and so it may be written as a map on the 2-torus:
\begin{eqnarray}
\label{mapj}
& \vartheta _{n+1} =\vartheta _n + J_n & \qquad\qquad {\rm mod}\;   (2\pi), \nonumber\\
%& J_{n+1} =J_n +q^2 \epsilon f_j(\vartheta,k) + 2\pi \Omega q &  \qquad\qquad {\rm mod}\;   (2\pi).
& J_{n+1} =J_n - \epsilon q^2 
\dot \omega_j(\vartheta,k) + 2\pi \Omega q &  \qquad\qquad {\rm mod}\;   (2\pi). 
\end{eqnarray}
In case $q=1$, this map reduces to one which was introduced in \cite{FGR022}, in order to explain the QAMs that had been experimentally observed near principal resonances. For
$q>1$, it has $q$ different versions, labeled by $l=1,\ldots,q$. Like in the case $q=1$, the stable periodic orbits of each of these versions  are expected to give rise to QAMs. Indeed,  each stable periodic orbit of map (\ref{mapj}) corresponds to a stable accelerating orbit of map (\ref{map0}), because the difference between momentum $I_n$ and momentum $J_n$ linearly increases with time. More precisely,
let
$(\vartheta _0, J_0)$ be initial conditions for  a periodic orbit of period $s$ and winding number $r/s$.  
%That means, the increment of $J$ after time $s$ (measured in the number of kicks), multiples of $2\pi$ %included, is $2\pi r$. 
The increment of $J$ after time $ns$ (measured in the number of kicks) is 
$2\pi r n$; therefore,  the increment of the original momentum variable is: 
\begin{eqnarray}
\label{ist}
I _ {sn} - I_0\;=\;a _I s n\;\;\;\;,\;\;\;
a _I =\frac {2\pi}{q} \left(
\frac r{qs} - \Omega \right)\;,
\end{eqnarray}
 with $I _0  =  (J _0 -\varrho)/q^2$.
This formula (\ref{ist}) yields  the acceleration of a stable orbit of the pseudoclassical dynamics (\ref{map0}), and it is precisely such orbits that may give rise to QAMs in the vicinity of resonances of arbitrary order.
As a matter of fact,  numerical 
simulations reveal QAMs near higher order resonances, in correspondence with
periodic orbits of maps (\ref{mapj}). 
%In fig. (\ref{phsp-q2p3-q7p11-q15p22}) 
%examples of phase space of maps
%(\ref{mapj}) are shown in parameter regimes in which QAMs are present.\\
In sect. \ref{exper}, we explain how the analysis of the stable periodic orbits of maps (\ref{mapj}) 
may help to resolve the complex pattern of QAMs presented in fig.\ref{pac-tot}\\
Thanks to (\ref{recipe}) and (\ref{clm}), the physical momentum $N$ is related to $I$ by $N= q I /\epsilon +j$; therefore, the physical acceleration is given by:
\begin{equation}
\label{acc}
a=\frac {2\pi}{\epsilon }\left(  \frac r{qs} - \Omega \right).
\end{equation}

Although the analytical derivation of the maps  is based on the resonant Hamiltonian, 
which is known in closed form only for $q=1,2$, the practical use of (\ref{mapj}) only 
requires the resonant eigenvalues, which can be easily computed 
by a  numerical diagonalization of a $q\times q$ matrix.

 In fig.\ref{phsp-q2p3-q7p11-q15p22}
examples of phase space of maps 
(\ref{mapj}) are shown for $q=2$ (a,b), $q=7$ (c) 
and $q=15$ (d), in parameter regimes in which QAMs are present. The plotted 
periodic orbits correspond to some of the modes shown in fig.\ref{pac-tot}.   
 For instance, in fig.\ref{phsp-q2p3-q7p11-q15p22} (a) 
 the stability island of a fixed point of one of the maps (\ref{mapq2}) is plotted 
 for $\tau /2\pi =1.455$; this fixed point corresponds to the huge mode on the left side of 
 fig.\ref{pac-tot}. A distribution of phase-space points, which initially fall inside the stability island, 
 describes an ensemble of atoms generating the QAM.
 
 Classical structures, like stability islands, may affect the quantum system only if their size is comparable with the effective Planck constant $\epsilon$. 
  The map (among the $q$ set of eq. (\ref{mapj})), that crucially contributes 
  in determining the observed QAMs, is  
 generally the one with the widest bandwidth.

\subsection {Special values of quasi-momenta.}
%\label{specqm}

A QAM arises when the initial wave packet is centered in momentum around $N_0$,
 related to $I _0$ by
\begin{equation}
\label{n0}
N _ 0 = q\frac {I _0} {\epsilon }+j = \frac 1{q\epsilon}( J_0+2\pi n) +q \alpha _j -
\frac {1}{\epsilon}\left( \pi\Omega+\tau\beta -2\pi p\bro /q\right)+j
\end{equation}
with $n\in {\mathbb Z}$.
As in the case for main resonances, we expect that the modes will be especially pronounced when quasimomentum is fine tuned: in view of (\ref{n0}) such optimal values of $\beta$ are determined by the condition:
\begin{equation}
\label{beta}
\beta _\nu =-\frac {\epsilon}{\tau} (N_0 -j-q \alpha _j +
\bro )+\frac {J_0+2\pi m}{q\tau}-\frac {\eta}{2}+\bro  \qquad {\rm mod}\; (1),
\end{equation}
with $\bro=\frac {\nu}{p} +\frac q2$ and $\nu =0,1,..,p-1$. A wave packet initially localized in $N_0 +
\beta_\nu$ will be mostly captured inside a QAM;  indeed in this case, the 
overlap between the stability island and the initial wave packet is maximal. 
 Formula (\ref{beta}) is a generalization of the result derived for $q=1$ in \cite{FGR022} 
  and  
experimentally verified in \cite{S};  it reduces to the expression in \cite{FGR022}  for $\alpha_0 =0$ (see appendix \ref{appq2-sp}).

\begin{figure}
  \includegraphics[width=8cm,angle=0]{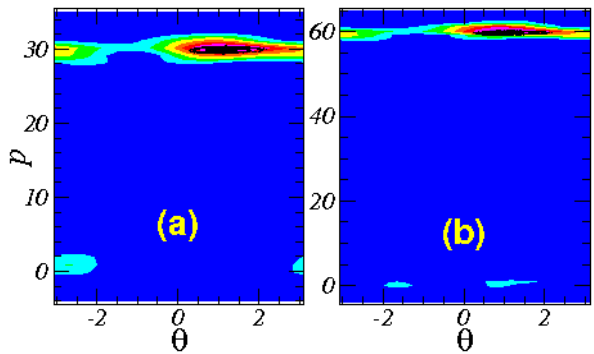}
  \caption{ Contour plots at times $t=100$ (a) and $t=200$ (b) 
  of the Husimi distribution of the wave packet 
  of a $\beta$-rotor with $\beta =0.1672$, given by (\ref{betaq2}) 
  with $j=1$ and $\nu =0$. The rotor is 
   initially prepared in a coherent state centered in the $(r,s)=(1,1)$ fixed point of 
   fig.\ref{phsp-q2p3-q7p11-q15p22}(a). 
  The black spots in the centers of the contours are an 
ensemble of classical phase points, initially distributed in a circle of area $\sim \epsilon$ centered at the 
mode. They evolve according to the $\epsilon$-classical dynamics (\ref{mapiq2}) with
 $\varrho =-0.5709$. The 
other parameter values are 
 $k=1, \tau =1.455*2\pi 
  (\epsilon =-0.2827)$ and $g=0.0386$.}
\label{hus-st}
\end{figure}

This picture 
is confirmed by  fig.\ref{hus-st}, in which the quantum phase-space 
evolution of a $\beta$-rotor, with a quasi-momentum
 given by (\ref{beta}), and the pseudoclassical motion are compared. The initial state of the 
 rotor is a coherent wave packet centered in the $(r,s)=(1,1)$ fixed point, plotted in 
 fig.\ref{phsp-q2p3-q7p11-q15p22}(a), corresponding to the 
  $\epsilon$-classical accelerator mode on the left part of fig.\ref{pac-tot}, in the vicinity of 
  $q=2$ resonance. The mode moves with an acceleration equal to 0.2988, 
  according (\ref{acc}).

\subsection{Validity of the Pseudo-Classical Description.}
\label{meaning}

We now come back to the meaning of the pseudo classical description, as 
  ``pseudoclassical" dynamics (\ref{map0}) still explicitly retains the ``Planck constant" $\epsilon$.
In the case when $q=1$, there is a  single resonant eigenvalue,
given by $\omega_0(\vartheta,k)=k\cos(\vartheta)$, so the pseudoclassical
dynamics (\ref{map0}) has a well defined limit
for $\epsilon\to 0$, $k\to\infty$, 
% $k\epsilon\to\tilde k<\infty$. 
$k\epsilon\to\tilde k$ with $|\tilde k|<\infty$.
This limit dynamics was discovered and analyzed in \cite{FGR021,FGR022}.
This is no longer true when $q>1$ and then the relation between the band dynamics and the ``pseudoclassical" dynamics (\ref{map0}) is less transparent. The quantum band-dynamics is still, formally, the quantization of the classical kicked dynamics
(\ref{map0}) using $\epsilon$ as the Planck constant. Nevertheless, the latter dynamics
contains the ``Planck constant" $\epsilon$ in crucial ways,
which preclude existence of a limit for $\epsilon\to 0$.
To see this, note that $\omega_j(\vartheta, k)$ depends on its arguments only through the real variables $u=k\sin(\vartheta/q), v=k\cos(\vartheta/q)$ (cf. the form of the resonant evolution  (\ref{evsp3}) in  
appendix \ref{exres})
%section (\ref{resspin}) \footnote{vedi la nota in quella sez.; occorre inserire l'espressione esplicita}), 
that is,
$\omega_j(\vartheta,k)=G(u,v)$, where $G$ is a smooth oscillatory function, independent of $k$.
 Hence,
\begin{equation}
\label{osc}
\epsilon \dot \omega_j (\vartheta ,k)
% \epsilon\;\partial_{\vartheta}\omega_j(\vartheta,k)
\;=\;\frac{\epsilon k}{q}\;
\left\{\cos(\vartheta/q)\partial_{u}G-\sin(\vartheta/q)\partial_{v}G\right\}
\end{equation}
Existence of a limit demands $\epsilon k\to\tilde k$; but then, except in the trivial case
$\tilde k=0$, the arguments of the $G$ functions in (\ref{osc}) diverge and so (\ref{osc}) appears to oscillate faster and faster as $\epsilon \to 0$, without a well-defined limit.\\
Nonexistence of a pseudoclassical  limit for the quantum dynamics 
was established in  \cite{GR08},
by a stationary phase approach, with no recourse to the band formalism. It was nonetheless pointed out that, despite
absence of such a limit, QAMs may be associated to certain rays,
which do correspond to trajectories of some formally classical maps. The meaning of the latter maps is, at most, that of providing local phase space descriptions, near QAMs.
Similar remarks apply in the case of the pseudoclassical maps (\ref{map0}).\\
It is worth recalling that maps (\ref{map0}) were derived  from an ansatz, which would be  optimally justified  if  the effective Hamiltonian of kicked dynamics could be replaced by the sum of the free and of the kicking Hamiltonians (sect. \ref{sodec}). This approximation is obviously  invalid in a global sense, yet, in ``spinless" cases,  it is known to work remarkably well near stable fixed points \cite{SFGR05}; indeed, in the KR case
it yields a pendulum Hamiltonian, which provides a good  description of the motion near the stable fixed point of the Standard Map.
This may be seen as a  qualitative justification for the use of maps (\ref{map0}), if restricted to the search of QAMs.

%\subsection {Analysis of QAMs.}
%\label{accqam}

\subsection {Case $q=2$.}
%\subsection {Case $q=2$ ($\tau ^{\rm res}=\pi p$) and $V(\theta)=k\cos \theta$.}
%\label{q2}
While the expressions in subsect. \ref{pcmaps} 
are quite general, we may accomplish a detailed analysis when $q=2$ and
 $V(\theta)=k \cos( \theta)$. In such a case
the eigenvalues $\omega _j (\vartheta , k )$  ($j=0,1$) of the
resonant Hamiltonian can be written down explicitly (see appendix \ref{appq2}):
\begin{equation}
\label{eigq2}
\omega _j (\vartheta ;k)=-\aap \left[ \frac {\pi}4+(-1)^j \arccos \left(
\frac {\cos (k \cos(\vartheta /2)}{\sqrt 2}\right) \right],
\end{equation}
with $\aap =(-1)^{\frac {p+1}{2}}$.
Therefore, our theory produces two maps (\ref{mapj}), which take the form:
\begin{eqnarray}
\label{mapq2}
& \vartheta _{t+1} =\vartheta _t +J_t & \quad\quad\quad {\rm mod}\; (2\pi),   \nonumber\\
& J_{t+1} =J_t + 4\pi\Omega - 2(-1)^j\aap \tilde k
\sin \left( \frac {\vartheta _{t+1}}2\right)
\frac {\sin \left( k \cos \left( \frac {\vartheta _{t+1}}2
\right) \right)}{\sqrt {1+\sin ^2 \left(
k \cos \left( \frac {\vartheta _{t+1}}2
\right)\right)}}&
\quad\quad\quad {\rm mod} \; (2\pi).
\end{eqnarray}

Going back to the time-dependent form the maps are written as
%From (\ref{map0}) we get the time-dependent maps on the cylinder:
\begin{eqnarray}
\label{mapiq2}
& \vartheta _{t+1} =\vartheta _t +4I_t  +4\pi\Omega t +\varrho
 & \quad\quad\quad {\rm mod}\; (2\pi),   \nonumber\\
& I_{t+1} =I_t  - (-1)^j\aap \frac {\tilde k}{2}
\sin \left( \frac {\vartheta _{t+1}}2\right)
\frac {\sin \left( k \cos \left( \frac {\vartheta _{t+1}}2
\right) \right)}{\sqrt {1+\sin ^2 \left(
k \cos \left( \frac {\vartheta _{t+1}}2
\right)\right)}}&
\quad\quad\quad {\rm mod} \; (2\pi),
\end{eqnarray}
with $\varrho =2\left( \epsilon\delta_{j,1}
 +\pi\Omega +\tau\beta -\pi p\beta_0 \right)$  and where we have used
 $\alpha _j=-\frac 12 \delta_{j,1}$  (see appendix \ref{appq2-sp}). 

 We remark that in the case $q=2$, avoided crossing between the 
 eigenvalues (\ref{eigq2}) are absent for an arbitrary value of $k$. 

\begin{figure}
  \includegraphics[width=8cm,angle=0]{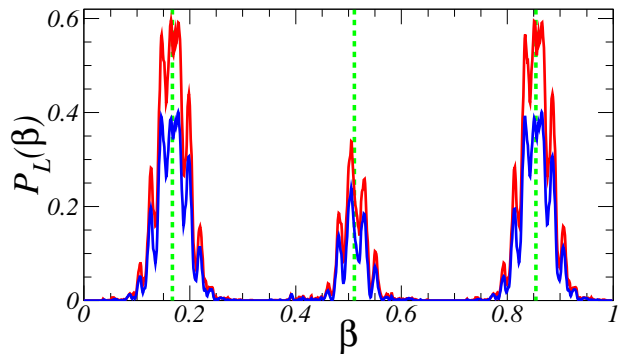}
  \caption{ Probability inside a box of extension equal to $L=6\simeq \Delta J q/|\epsilon |$ in momentum, moving according to (\ref{acc}) for $p=3, q=2, (r,s)=(1,1)$ and $\epsilon =-0.2828$,  as a function of quasimomentum $\beta$ of the $\beta$-rotor.  $\Delta J$ is the size of the island in $J$. Dashed vertical lines refer to special values of quasimomenta, given by formula (\ref{betaq2}) with 
$\bro =\nu/3$, $\nu =0,1,2$, $N_0=0, m=0$. 
The probability is shown at time $t=100$ (red) and $t=200$ (blue). 
The parameter values are the same as in fig.\ref{pac-tot}.}
\label{qmom}
\end{figure}

We may also check the selection criterion for quasimomenta, that in the present case assumes the form:
\begin{equation}
\label{betaq2}
\beta _{j, \nu}=-\frac {\epsilon}{\tau} (N_0 +\bro )+\frac {J_0+2\pi n}{q\tau}-\frac {\eta}{2}+
\bro  \qquad {\rm mod}\; (1),
\end{equation}
with $\beta _0$ given by $(iii)$ in sect. \ref{back} and $\nu=0,...,p-1$. 
 A scan over possible $\beta$ values reveals that indeed QAMs are greatly enhanced around the values predicted by (\ref{beta}):   this is confirmed by  fig.\ref{qmom}, in which the momentum 
 probability transferred to the mode is shown as a function of $\beta$.

%%%%%%

\section{Mode spectroscopy and connections with cold atom experiments.}
\label{exper}

\subsection{Farey ordering of QAMs near a fixed resonance.}

We now elucidate how our findings apply to inspection of density plots like the one illustrated in 
fig.\ref{pac-tot}. We point out that such a picture is of direct physical significance, since typical experimental protocols maintain $k$ and $g$ fixed, while performing a scan on the pulse period $\tau$. Such a scan, in the present context, has to be carried out around a resonant value, namely
  $\tau (\epsilon ) =2\pi p /q +\epsilon$. Density plots of momentum distribution disclose the presence of QAMs, since after a fixed number of kicks their momentum is linearly related to the acceleration (\ref{acc}): $a$ depends on $\epsilon$, through the ``bare" winding number $q\Omega$
%Fig.\ref{pac-tot} (distribuzione momento fig.1)
\begin{equation}
\label{ep}
q\Omega (\epsilon )  =\frac q{2\pi} g\left( 2\pi \frac pq +\epsilon \right)^2
\end{equation}
and on the ``dressed" winding number of the pseudo classical map $r/s$, which individuates the mode. We denote by $\Omega^*$ the resonant ($\epsilon=0$) value of the ``bare" winding number (notice that for $\epsilon=0$ the maps correspond to pure
rotation in $J$):
\begin{equation}
\label{oms}
 \Omega ^ * \equiv q\Omega (0)=2\pi \frac {p^2}q g,
\end{equation}
which is independent of the mapping index $j$.  Formula (\ref{oms}) is a generalization 
of the analogous result found for $q=1$ \cite{GRF05,all05}. 

As analyzed in \cite{GRF05,all05} for  
principal resonances, the parameter space of map (\ref{mapj}) is 
characterized by the presence of regions (Arnol'd tongues), in which stable periodic 
orbits exist. 
Close to resonances we expect that mode-locking structure of the pseudo classical maps singles out modes whose winding number provide rational approximants to $\Omega^*$: at the same time fat tongues are associated to small $s$ values, so the corresponding modes should be more clearly detectable. This is the physical motivation underlying Farey organization of observed modes: whenever we observe two modes labelled by winding numbers $r_1/s_1$ and $r_2/s_2$ ($r_1/s_1 < \Omega^* < r_2/s_2$), the fraction with smallest denominator bracketed by the winding pair is the Farey mediant $(r_1+r_2)/(s_1+s_2)$.

We can now analyze in more detail fig.\ref{pac-tot}, which represents
a numerical simulation of experimental momentum distribution after $t=100$ vs $\tau$, 
which assumes values 
 around a second order resonance, namely $\tau ^{\rm res}/2\pi =p/q=3/2$.
 All parameters are chosen to be accessible to experiments and the initial atomic distribution
  reproduces that employed in \cite{Ox991,Ox992,Ox995,Ox994}:  
  $g=0.0386$, $k=1$  and the initial state is a mixture of 100 plane waves sampled from a gaussian distribution of momenta with FWHM $\sim 9$.

Full lines in the figure delineate momentum profiles consistent with the acceleration (\ref{acc}),
with $(r,s)$ given by winding number $r/s$ of corresponding stable periodic orbits of maps (\ref{mapj}).

\begin{figure}
  \includegraphics[width=8cm,angle=0]{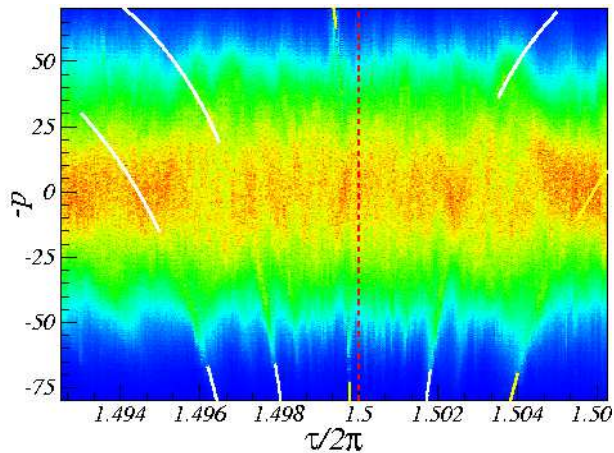}
  \caption{ Enlargement of fig.\ref{pac-tot} in the region $1.49\leq\tau/2\pi\leq 1.50625$, around the resonance $\tau_{res} =3\pi$ $(p/q=3/2)$. Momentum distributions is calculated after
$t=200$ kicks. Full lines show the theoretical curves (\ref{acc}): the yellow ones refer to principal convergents of $\Omega ^*$. Starting from the left, the modes correspond to the stable periodic orbits of maps (\ref{mapq2}) with:
$(r,s)=(14,13), (25,23), (12,11), (23,21)$ and $(11,10)$.}
\label{pac-q2p3}
\end{figure}

 The value of $\Omega ^*$ and the first few rational approximants (obtained upon successive truncation of the continued fraction expansion),  corresponding to detectable modes, are:
\begin{eqnarray}
\label{ostar1}
& & \Omega ^* \simeq 1.0913893 = 1 + [10,1,16,3,3,...]\nonumber\\
& & \frac rs = 1; \; \frac {11}{10};\;  \frac {12}{11}; ...
\end{eqnarray}
The first one $(r,s)=(1,1)$ is shown with yellow full line on the left of fig.\ref{pac-tot} 
and the stability island of the corresponding fixed point is shown 
in fig.\ref{phsp-q2p3-q7p11-q15p22} (a).
The second and third are marked by full yellow lines in fig.\ref{pac-q2p3},
which is an enlargement of fig.\ref{pac-tot} in the region $1.49\leq \tau/2\pi\leq 1.50625$, calculated for time $t=200$.
Farey organization is exemplified by the appearance of the $(23,21)$ QAM, whose winding number is the Farey composition of the $(11,10)$ and the $(12,11)$ modes;  the correspondent stable periodic 
orbit is plotted in fig.\ref{phsp-q2p3-q7p11-q15p22}(b). 
Through Farey composition law we may also identify observed modes to the right of $\tau_{res}$, as shown in  fig.\ref{pac-q2p3}.

\subsection{Visibility of resonances of different order.}

\begin{figure}
  \includegraphics[width=8cm,angle=0]{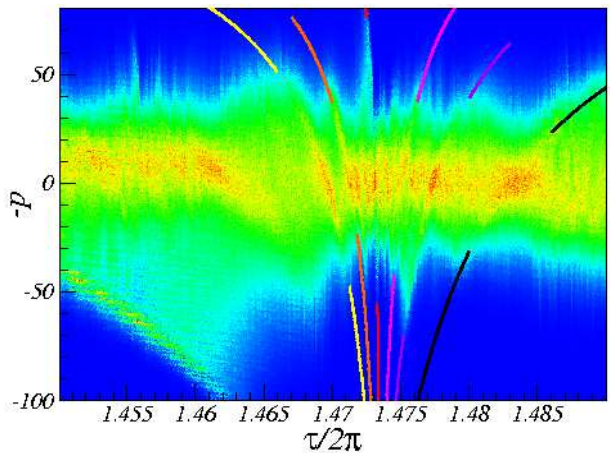}
  \caption{ Enlargement of fig.\ref{pac-tot} in the region $1.45\leq\tau/2\pi\leq 1.49$. The momentum distributions is calculated after
$t=200$ kicks. Full lines show the theoretical curves (\ref{acc}); each color refer to a different quantum resonance, namely different values of $p/q$. Starting from the left the modes correspond to:
$p/q=3/2$ (theoretical curve not shown), $31/21, 59/40, 28/19, 53/36, 25/17$ and $22/15$. These 
resonances $p_n/q_n$ 
lead to almost integer $\Omega^*$ and they are exacted from the sequence of Farey 
fractions obtained starting from  $p_0/q_0=1/1$ and $p_1/q_1=3/2$: 
$p_8/q_8=22/15,\; \Omega ^*_8\simeq 7.82566 =8-[5,1,2,1,...],\; (r,s)=(8,1)$ (shown in black); 
$ p_9/q_9=25/17, \; \Omega ^*_9\simeq 8.9165791=9-[11,1,78,...], \; (r,s)=(9,1)$ (in purple); 
$p_{10}/q_{10}=28/19; \; \Omega ^*_{10}\simeq10.0075= 10+[131,...], 
\; (r,s)=(10,1)$ (in red); $p_{11}/q_{11}=31/21,\;  \Omega ^*_{11}\simeq11.098678=11+[10,7,2,6,1,...], 
 \; (r,s)=(11,1)$ (in yellow).  Further modes are shown in between the mentioned ones: 
 $p/q=53/36 =25/17\oplus 28/19$ with $\Omega ^* \simeq 18.924152 = 19-[13,5,2,...]$ (shown in pink) and  $p/q=59/40 =28/19\oplus 31/21$ with $\Omega ^* \simeq 21.106256 = 21+[9,2,2,3,...]$
(shown in orange). The $(8,1)$-periodic orbit of the resonance $p_8/q_8=22/15$ is plotted 
in fig.\ref{phsp-q2p3-q7p11-q15p22}(d). 
}
\label{pac-qalto}
\end{figure}

The complexity of mode spectroscopy is further enhanced by the fact that, within 
some interval in $\tau$,  arbitrarily many different resonant values occur. As a matter of fact it is possible to recognize in
fig.\ref{pac-tot} modes coming from a wide set of resonances:  
besides $q=2$ also $q=7, \,15,\,17,\,21,\,36,\,40$ contribute QAMs in the selected range; 
this is shown in fig.\ref{pac-tot} for $q=7$ and in fig.\ref{pac-qalto} for the other resonances.  
No QAM with $q=13$ could be resolved in the range of fig.\ref{pac-tot}. 

Farey composition is still of some use in the identification of the resonances to which modes belong: for instance, the very large mode
on the right of the figure belongs to a QR between $p/q=3/2$ and $p/q=2/1$; applying Farey composition successively, we get the sequence $p/q=5/3, 8/5$ (outside the plotted range in $\tau$) 
and then $11/7$, to which the mode belongs.
The accumulation point of the resonance $p/q=11/7$ is $\Omega ^*\simeq 4.1923207 =4+[5,...]$. The mode shown in fig.\ref{pac-tot} 
 corresponds to the first principal convergent of $\Omega ^*$, i.e. to the fixed point 
 $(r,s)=(4,1)$,  shown in fig.\ref{phsp-q2p3-q7p11-q15p22}(c). The same occurs 
 for the modes near resonances of higher $q$, shown in fig.\ref{pac-q2p3}. 

We remark that a hierarchy in resonant fractions looks more cumbersome than the one considered for winding numbers, as for instance there does not seem to be any straightforward dependence on the size of $q$. Numerical data however suggest that detectable modes appears in 
the vicinity of resonances leading to almost integer $\Omega^*$, i.e. when 
the fractional part of $\Omega^*$ is 
 closer to the integers 0 or 1 than to their Farey mediant 1/2.  
 In these cases, the resonance may display a mode corresponding to a periodic orbit of period 1.  
 As shown in fig.\ref{pac-qalto}, 
this condition may be fulfilled for different $p/q$ values. Moreover, the 
absence of observable QAMs with $q=13$  in the range of fig.\ref{pac-tot}, even if $p/q=20/13$ resonance belongs to the plotted $\tau$ range, is consistent with this rough ``thumb" rule.  
Indeed $\Omega^* \simeq 7.4624908 = 7 + [2, 6, 6, ..]$, thus its fractional part is closer to 1/2 than to 0.

%%

%\section{Concluding remarks, prospectives and acknowledgements.}
\section{Summary.}

The quantum dynamics of quantum accelerator modes, experimentally  observed 
by exposing cold atoms to periodic kicks in the direction of the gravitational field, is 
theoretically described in terms of spinors, when the pulse period is close to a rational multiple 
of a characteristic time of the atoms (Talbot time). The reference model is 
a non-trivial variant of the well-known Kicked Rotator in an almost-resonant regime. 
If  the detuning of 
the kicking period to the resonant value  is assigned the role of the 
 Planck constant, the problem is shown to 
share similarities with the semiclassical limit of the particle  
dynamics in presence of spin-orbital coupling. The separation of the spinor and orbital 
degrees of freedom, is based on an ``adiabatic" assumption of Born-Oppenheimer type,  
valid for small detunings and for values of the parameters in which the QAMs manifest. 
In these parameter regimes,  a 
 description of some properties of the ``slow" orbital motion, by means of formally classical equations,  
is finally achieved.  Some results of a previously formulated ``pseudo-classical" 
theory \cite{FGR021}, restricted to QAMs near principal resonances, are 
extended to arbitrary higher order resonances. Potential applications to current experiments on 
cold atomic gases are proposed.

% possibili prospettive future?
%
%1) comprensione piu' approfondita del problema BO approx, ham. efficace e sua forma 
% approssimata vicino alle orbite periodiche stabili
%
%2) approccio che usa la condizione di risonanza con gravita' come punto di partenza
% (dana:Phys. Rev. E 76, 015201(R) (2007), lemarie); 
%in particolare $\Omega=r/s$.

L.R. acknowledges useful discussions with Shmuel Fishman.

%%%%
\appendix
%%%%%%%%%
\section{Spinor dynamics at exact resonance and without gravity.}
\label{exres}
The Floquet operator at exact resonance in absence of gravity $\eta=0$
is given by (cp. (\ref{fo})) :
\begin{equation}
\label{fores}
\hat U_{\rm res} =e^{-ikV(\hat \theta )} \cdot e^{-i\pi \frac pq (\hat N +\bro )^2},
\end{equation}
where $\bro=\nu/rp +rq/2$ with $\nu\in{\mathbb Z}$.
The evolution of spinor components $\phi _j \vt$ under
$\hat U_{\rm res}$ is given by \cite{cs}:
\begin{eqnarray}
\label{evsp1}
& & \bar {\phi} _j(\vartheta)=
\sum_{l =0}^{q-1}(e^{-ik\hat V})_{j-l}\; e^{-i\pi\frac pq (l+\bro )^2} {\phi} _{l} (\vartheta)\\
\label{evsp3}
& & (e^{-ik\hat V}) _{j-l} =
\langle j | e^{-i\frac {\vartheta}{q} \hat {\bf S}}
\fft e^{-ik\hat {V}(\theta )}\fft ^\dagger  e^{i\frac {\vartheta}{q} \hat {\bf S}} | l\rangle =
\frac {e^{-i(j-l)\frac {\vartheta}{q}}}{q}
\sum_{m =0}^{q-1} e^{-i\frac {2\pi}q m (j-l)} e^{-ik{\hat V}\left( \frac {\vartheta}{q}
+\frac {2\pi}{q}m \right)}.
\end{eqnarray}
$\fft$ is the Fourier transform in ${\mathbb C}^q$: $\langle j | \fft | l \rangle =e^{-i
\frac {2\pi}{q}jl}/\sqrt {q}$ and $\hat {\bf S}$ is the spin operator (\ref{bthspin}).
Hence from (\ref{evsp1}), the resonant Floquet operator is decomposable
 in spin propagators  $\fib\vt$, given by unitary matrices of rank
$q$; namely it acts as
$(\hat {U}_{\rm res} {\phi}) \vt={\fib} \vt \pmb {\phi} \vt$.

%%%
\section{Case $q=2$: resonant eigenvalues and eigenvectors.}
\label{appq2}

At primary 2nd order ($q=2$) resonances,  the resonant values of quasimomentum are
$\beta _0 =\nu/p$ with $\nu=0,1,..,p-1$. We choose $\beta_0 =0$.
From (\ref{fibfl1}) and (\ref{fibfl2}), denoting $v(\vartheta ;k) = k\cos(\vartheta /2)$ and $\aap =
(-1)^{\frac {p+1}{2}}$, we find
\begin{equation}
\label{fibra1}
 \fib _{k,p,2 , 0 } =
\left(\begin{array}{cc}
\cos (v(\vartheta ;k)) & \aap e^{i\frac \vartheta 2}\sin (v (\vartheta ;k) )\\
-i e^{-i\frac \vartheta 2}\sin (v (\vartheta ;k))  & i\aap  \cos (v(\vartheta ;k ) )
\end{array}\right).
\end{equation}

Matrix (\ref{fibra1}) can be written in terms of Pauli matrices as follows:
\begin{eqnarray}
\label{fibrapau1}
\fib _{k,p,2 , 0 } &=& e^{i\aap \frac {\pi}{4}}\left( \cos (\bar \omega (\vartheta ;k ))+i\;
\vec {\boldsymbol x}(\vartheta ;k) \cdot \vec {\boldsymbol \sigma }
\right), \\
\label{fibrapau1-1}
&=& e^{i\aap \frac {\pi}{4}}\left( \cos (\bar \omega (\vartheta ;k ))+i\sin (\bar \omega
(\vartheta ;k ))\; (\vec {\boldsymbol n}(\vartheta ;k) \cdot \vec {\boldsymbol \sigma )}
\right),\\
\label{eigvq2-omega}
& & \bar {\omega} (\vartheta ;k) = \arccos \left( \frac {\cos (k\cos (\vartheta /2))}{\sqrt 2}\right),
\end{eqnarray}
where the vector ${\boldsymbol x}(\vartheta ;k)\in {\mathbb R}^3$ has components:
\begin{eqnarray}
\label{componenti}
& & x_1(\vartheta ;k) = \aap \sin \left( \frac \vartheta 2 -\aap \frac \pi 4\right)\sin (v(\vartheta ;k)),
 \nonumber \\
 & & x_2(\vartheta ;k) = \sin \left( \frac \vartheta 2 +\aap \frac \pi 4\right)\sin (v(\vartheta ;k)),
 \nonumber \\
& & x_3(\vartheta ;k) = -\frac {\aap}{\sqrt 2}\cos (v(\vartheta ;k));
\end{eqnarray}
$\vec {\boldsymbol n} =\vec {\boldsymbol x}/
x$ is a unit vector in ${\mathbb R}^3$, and
$x(\vartheta ;k)= ||{\boldsymbol x}(\vartheta ;k) ||$.

Using a well-known formula, equation (\ref{fibrapau1}) may be written:
\begin{equation}
\label{fibrapau2}
\fib _{k,p,2 , 0 } =  e^{i\aap \frac {\pi}{4}} e^{i\bar \omega (\vartheta ;k )\; \vec {\boldsymbol n}
(\vartheta ;k )
\cdot \vec {\boldsymbol \sigma}},
\end{equation}
which directly yields to the Resonant Hamitonian for this case:
\begin{equation}
\label{hresq2}
{\bf  \hat H}^{\rm res}_{k,p,2,0}=-\left[ \aap \frac {\pi}{4}+\bar \omega  (\vartheta ;k )\; \vec {\boldsymbol n}
(\vartheta ;k )
\cdot \vec {\boldsymbol \sigma}\right] .
\end{equation}
As matrix $\vec {\boldsymbol n} \cdot \vec {\boldsymbol \sigma}$
has eigenvalues $\pm 1$, the eigenvalues of the resonant fiber are:
\begin{eqnarray}
\label{eigvq2}
& & \lambda ^{(j)}(\vartheta , k) =e^{i\aap \left[  \frac \pi 4 +(-1)^j \bar {\omega} (\vartheta ;k)\right]}
=e^{-i \omega _j (\vartheta ;k)} \\
\label{eigvq2-1}
& & \omega _j (\vartheta ;k) = - \aap \left[ \frac {\pi}{4} +(-1)^j \bar {\omega} (\vartheta ;k)\right]
\qquad\qquad j=0,1.
\end{eqnarray}
Normalized eigenvectors are:
\begin{eqnarray}
\label{eigvecq2}
& &{\pmb \varphi} _j (0,\vartheta) =
e^{i\gamma _j(\vartheta)} \frac {\sin (v(\vartheta ;k))}{\sqrt {a_j(\vartheta ;k )}};\\
\label{eigvecq2a}
& & {\pmb \varphi} _j (1,\vartheta) = e^{i\gamma _j(\vartheta)} e^{-i\frac {\vartheta}{2}}\frac {i-\aap }{2}\frac {b_j(\vartheta ;k)}{\sqrt {a_j(\vartheta ;k )}} ; \\
\label{eigvecq21}
& & a_j(\vartheta ;k ) = 1+\sin ^2 (v(\vartheta ;k ))+(-1)^j\cos (v(\vartheta ;k ))\sqrt
{1+\sin^2 (v(\vartheta ;k ))}, \\
\label{eigvecq22}
& &  b_j(\vartheta ;k)=\cos (v(\vartheta ;k))+(-1)^j \sqrt {1+\sin^2 (v(\vartheta ;k ))}.
\end{eqnarray}
The $\gamma _j (\vartheta)$ are arbitrary phases. For $n_0 \pi \leq k < (n_0+1)\pi$ ($n_0 \in
{\mathbb N}$), $a_{1} (\vartheta ;k )$ has a finite set of zeros in $\{ \vartheta _n =\pm 2
\arccos (\pi n /k), 0\leq n \leq n_0 \}$.
Discontinuities in the zeros of $a_{1}
(\vartheta ;k )$ may be removed by appropriate choices of $\gamma _{1}$. For $k<\pi$
 and $-\pi < \vartheta \leq \pi$, one may choose, e.g. $\gamma _j (\vartheta) =\frac {\vartheta}{2}
 \delta _{j,1}$.

 For increasing values of the kicking strength $k$, eigenvalues (\ref{eigvq2-1}) display thicker and
 thicker oscillations. An example is shown in fig.\ref{levels} (a), for $k=1,3,5$. Nevertheless,
 contrary to higher values of $q$, in case $q=2$, the two eigenvalues (\ref{eigvq2-1})
 neither  cross
 nor become closer that a minimal gap, equal to $\pi /2$, for arbitrary high values of $k$.
 As a matter of fact, the band width of each eigenvalue,
 defined as $B_j (k)=|\bar \omega (\vartheta _{max};k)-\bar \omega (\vartheta _{min};k)|$,
 with $\vartheta _{max}$ and $\vartheta _{min}$ absolute maximum and minimum points
 in $[0, 2\pi[$, does not exceed $\pi /2$. For $k<\pi$, $B_j (k)$
 is an increasing function of $k$ equal to $B_j (k)=|\bar \omega (0;k)-\bar \omega (\pi;k)|=
 |\arccos (\cos (k\cos(\vartheta /2) /\sqrt 2))-\pi /4|$; for $k>\pi$,
 $B_j (k)=|\bar \omega (\vartheta_1 ;k)-\bar \omega (\pi ;k)|=
 \pi /2$.
 
 For $\beta_0 =\nu /p$, the same results hold, apart for a constant phase factor in (\ref{fibra1}) 
 and $m_p$ replaced by $-m_p$ for $\nu$ odd.

%%%
\subsection{Vector and Scalar potential.}
\label{appq2-sp}

The vector and scalar potentials can be explicitly computed, using
analytical expressions of the eigenvectors of resonant fiber (and choosing  phases
$\gamma _j (\vartheta) =\frac {\vartheta}{2}\delta _{j,1}$).
First one computes:
\begin{eqnarray}
&& S_{jj} \vt =\sum_{l=0}^1 l |\langle l |\pmb \varphi_j \vt \rangle |^2
= \frac 12 \frac {(b_j )^2}{a_j},  \nonumber \\
&& S^{''}_{jj}\vt =\sum_{l=0}^1 l^2 |\langle l | \pmb \varphi_j \vt \rangle |^2=
 S_{jj}, \nonumber \\
&&  S^{'}_{jj}\vt =\sum_{l=0}^1 l \langle \pmb \varphi_j \vt   | l \rangle
 \langle l | \pmb {\dot\varphi}_j \vt \rangle=\frac 1{2a_j}
 \left( b_j\dot {b}_{j}-\frac {1}{2a_j}\dot {a_j}(b_j)^2
 +\frac i2 (b_j)^2 (\delta_{j,1}-1)\right),\nonumber \\
 && i \langle \pmb{\varphi}_j \vt |  \pmb {\dot {\varphi}}_j \vt \rangle=
 -\dot \gamma _j \vt + \frac 14 \frac {(b_j )^2}{a_j}, \nonumber \\
 && \langle \pmb{\dot {\varphi}}_j \vt |  \pmb{\dot \varphi}_j \vt \rangle=
 \frac 1{a_j}\left( (\dot v )^2 \cos^2 v
 +\frac 12  (\dot {b}_{j})^2+\frac 18({b}_{j})^2
 -\frac 14 \frac {(\dot {a}_{j})^2}{{a}_{j}}\right)
 +\frac 14 \delta_{j,1}\left( 1-\frac {({b}_{j})^2 }{{a}_{j}}\right),\nonumber
\end{eqnarray}
and then the vector and scalar potentials are given by:
\begin{eqnarray}
&& {\mathcal A}_j \vt\equiv \alpha _j =-\frac {1}{2}\delta _{j,1},\\
&&{\mathcal B}_j\vt =
\frac 4{a_j}\left( (\dot v )^2 \cos^2 v
 +\frac 12  (\dot {b}_{j})^2
 -\frac 14 \frac {(\dot {a}_{j})^2}{{a}_{j}}\right) =
 \frac {2(\dot {v} )^2}{(1+\sin ^2 v )^2},
\end{eqnarray}
with $v\equiv v(\vartheta ,k)=k\cos (\vartheta /2)$.

For $q=1$ the resonant fiber is a matrix with a single element $\exp (-iv(\vartheta ,k))$, with
eigenfunction $\varphi _{0}\vt =\exp (in\vartheta)/\sqrt {2\pi}$
($n\in {\mathbb Z}$); $S_{00}=0$ and therefore
(\ref{ab1}) it yields ${\mathcal A}_0 \vt\equiv \alpha_0=-n$ with $n\in {\mathbb Z}$ and ${\mathcal B}_0\vt =0$.

%%%%%%

%%%%%%%%%%%%%%%%%%%%%%

\end{document}